\documentclass[final,1p,times]{elsarticle} 
\usepackage{graphicx} 
\usepackage{amssymb} 
\usepackage{amsthm} 
\usepackage{lineno} 

\journal{Nuclear Physics A} 

\def\simge{\mathrel{
    \rlap{\raise 0.511ex \hbox{$>$}}{\lower 0.511ex \hbox{$\sim$}}}}
\def\simle{\mathrel{
    \rlap{\raise 0.511ex \hbox{$<$}}{\lower 0.511ex \hbox{$\sim$}}}}

\begin{document} 

\begin{frontmatter} 


\title{Quarkonium in a viscous QGP}

\author{Adrian Dumitru}

\address{Department of Natural Sciences, Baruch College, 
A-506, 17~Lexington Ave, New York, NY 10010, USA}
\address{RIKEN-BNL Research Center, Brookhaven National Lab,
Upton, NY 11973, USA}

\begin{abstract} 
I discuss viscosity corrections to thermal effects on the static
QCD potential within hard-thermal loop resummed perturbation theory
and for a strongly coupled, large-$N_c$ conformal field
theory dual to five-dimensional Gauss-Bonnet gravity. I also present
model predictions for quarkonium binding energies in the deconfined
phase and for suppression of $R_{AA}(\Upsilon \to e^+e^-)$.
\end{abstract} 

\end{frontmatter} 



The potential between a very heavy $Q\bar{Q}$ pair in a color-singlet
state is approximately given by Coulomb attraction at short distances and
linear confinement at large separation~\cite{Eichten:1979ms},
\begin{equation}  \label{Cornell_T=0}
V_{Q\bar{Q}}(r) = - \frac{\alpha_s C_F}{r} + \sigma r~.
\end{equation}
Here, $\sigma\approx 1$~GeV/fm denotes the string tension in SU(3)
gauge theory. (The string is in fact screened in full QCD with
dynamical fermions due to ``string breaking'' at $r\simge 1/\Lambda_{\rm
  QCD}$.)  At $m_Q\to\infty$, bound states have small radii and hence the
Coulomb attraction dominates. If $\alpha\equiv\alpha_s C_F\ll 1$, the
binding energy of the ground state $|E_{\rm bind}| \sim
\alpha^2 m_Q$ is much smaller than the quark mass $m_Q$
and hence the velocity of the quarks in the bound state is small,
$v\ll 1$. Furthermore, the Bohr radius $a_0 \sim 1/(\alpha m_Q) \gg
1/m_Q$; the bound quarks are therefore not localized in a region on
the order of their Compton wavelength. These observations suggest
that to first approximation quarkonium states can be understood from
non-relativistic potential models~\cite{Lucha:1991vn} such as
(\ref{Cornell_T=0}). A framework for systematic improvements of this
simple picture is offered by an effective field theory (potential
non-relativistic QCD - pNRQCD) obtained from QCD by
integrating out modes above the scales $m_Q$ and then $m_Q v$,
respectively \cite{Brambilla:2004jw}.

At high temperature, the deconfined phase of QCD exhibits screening of
static color-electric fields~\cite{GPY}. Hence, quarkonium states
should dissociate once the Bohr radius exceeds the screening
length~\cite{KMS}. In recent years, a big effort has been made by
various groups to test the validity of potential models at finite
temperature, to compute thermal modifications of the potential, and to
obtain quarkonium spectral functions and meson current correlators via
first-principle QCD calculations performed numerically on a
lattice. We refer to ref.~\cite{Mocsy:2008eg} for a summary and
review. A qualitatively new contribution to the static potential which
arises at finite temperature is the imaginary part due to Landau
damping of the static gluon exchanged by the heavy
quarks~\cite{Laine:2006ns}.

Here, we focus on non-equilibrium effects in a plasma which exhibits a
local anisotropy. This arises in heavy-ion collisions due to
anisotropic hydrodynamic expansion of a plasma with non-vanishing
shear viscosity. The phase-space distribution of thermal excitations is
given by
\begin{equation}
f({\bf p}) = f_{\rm iso}(p) \left[ 1-\xi \frac{({\bf p}\cdot{\bf n})^2}
{2pT} \left( 1\pm f_{\rm iso}(p)\right)
\right]~.  \label{eq:f_aniso}
\end{equation}
$f_{\rm iso}(p)$ is either a Bose distribution or a Fermi
distribution, respectively. The correction $\delta f$ to the
equilibrium distribution follows from viscous hydrodynamics / kinetic
theory for a fluid element expanding one-dimensionally along the
direction ${\bf n}$; the anisotropy parameter $\xi$ is proportional to
the ratio $\eta/s$ of shear viscosity to entropy density and to the
gradient of the flow velocity.

To derive the potential in a plasma described by the momentum
distribution~(\ref{eq:f_aniso}) one first computes the corresponding
retarded and symmetric ``hard thermal loop'' resummed gluon
propagators in the static limit. The one-gluon exchange potential
follows essentially from its Fourier transform. Its real part is given
by~\cite{Dumitru:2007hy,Dumitru:2009ni} 
\begin{eqnarray} \label{eq:anisoPotlin_xi}
V({\bf{r}}) &=& {V}_{\rm iso}(r)
\left(1+\xi {\cal F}(\hat{r},\theta) \right)~, \\ 
{\cal  F}(\hat{r},\theta) &\simeq& +\frac{\hat{r}}{6}+\frac{\hat{r}^2}{48}
+\frac{\hat{r}^2}{16}\cos(2\theta)+\cdots 
\end{eqnarray} 
Here, $\hat{r}\equiv r\,m_D$ with $m_D(T)$ the screening mass in an
isotropic medium, and ${V}_{\rm iso}(r)=-\frac{\alpha}{r}\exp(-\hat{r})$ is
the well-known Debye-screened Coulomb potential. The viscosity-dependent
correction in eq.~(\ref{eq:anisoPotlin_xi}) reduces thermal screening
effects as compared to an ideal ($\xi=0$) plasma.

The potential (\ref{eq:anisoPotlin_xi}) may fail to reproduce the
dominant $T$-dependence of the binding energies in the
phenomenologically relevant range $T/T_C = 1-3$, even for very large
quark mass. In this range, the ``interaction measure'' $(e-3p)/T^4$ in
SU(3)-YM (or QCD) is large. The free energy of a static $Q\bar{Q}$
pair at infinite separation behaves as~\cite{LattF}
\begin{equation}  \label{FQ_Latt}
F_\infty(T) \simeq \frac{a}{T} - bT~,
\end{equation}
with $a\approx 0.08$~GeV$^2$ a constant of dimension two and $b$ a
dimensionless number. The second term is usually identified with an
entropy contribution which should be removed. The first term, however,
corresponds to a non-vanishing $V_\infty(T)\sim a/T$ tied to the
presence of an additional dimensionful scale besides $T$. In fact, for
very small bound states, the temperature dependence of the {\em
  short-distance} potential is much smaller than that of the continuum
threshold $V_\infty(T)$~\cite{Dumitru:2009ni}. Note that the binding
energy of a quarkonium state is defined relative to the potential at
infinity: $E_{\rm bind}= \langle\Psi\left|\hat{H}-V_\infty\right|
\Psi\rangle - 2m_Q$.

\begin{figure}[ht]
\centering
\includegraphics[width=0.49\textwidth]{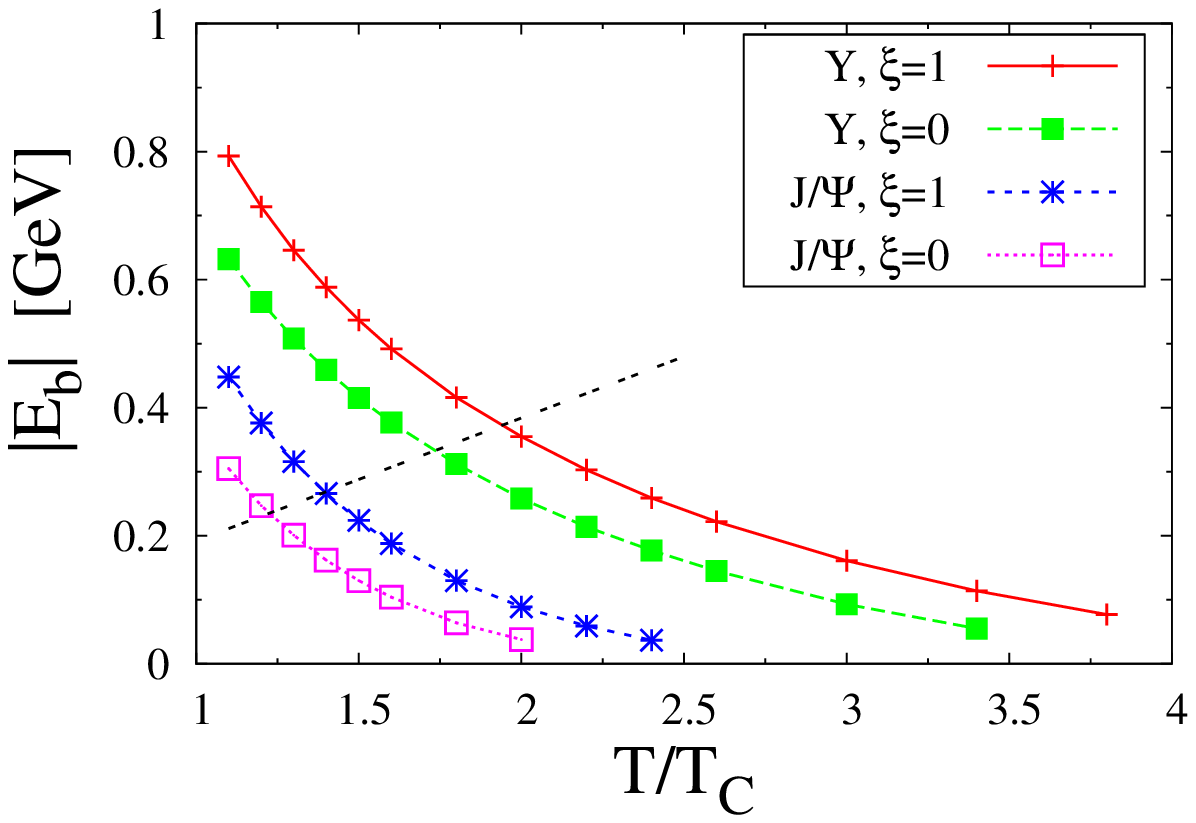}
\includegraphics[width=0.49\textwidth]{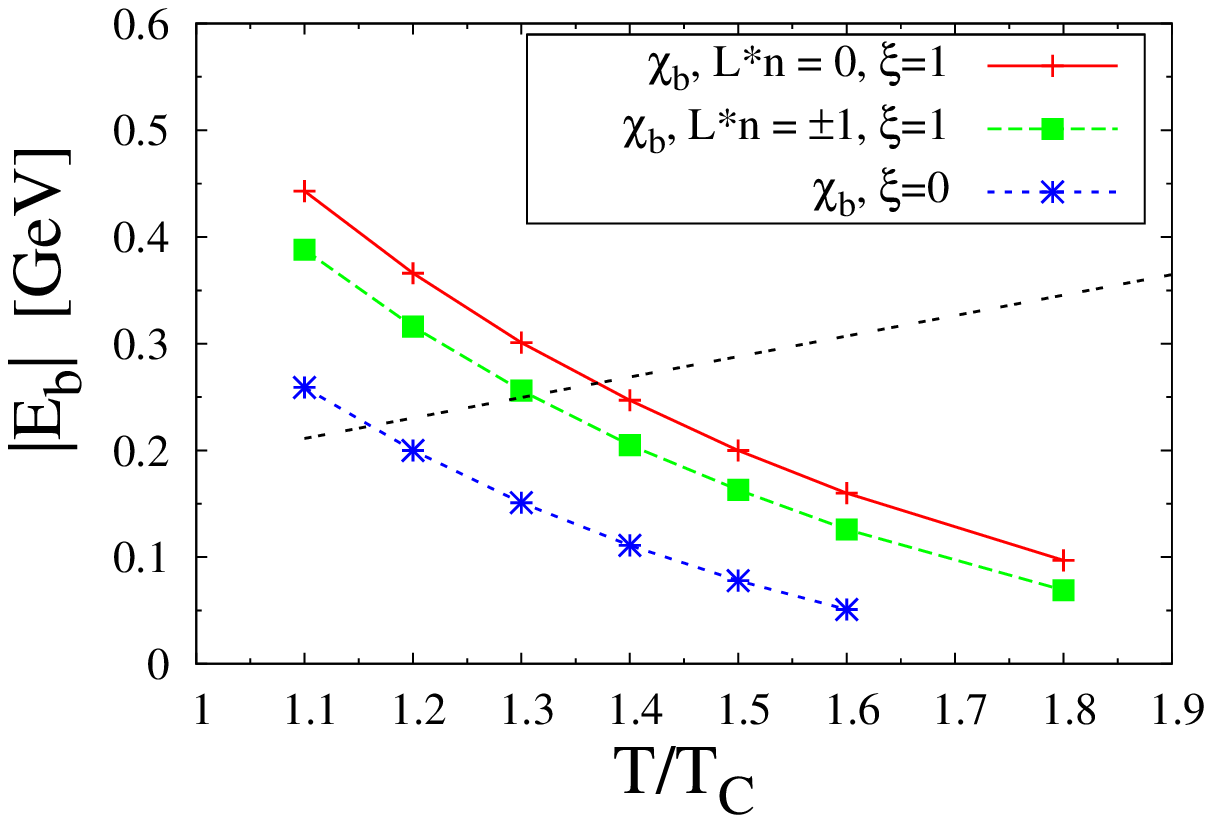}
\vspace*{-.4cm}
\caption[a]{Left: Binding energies for the 1S states of charmonium
  (lower curves) and bottomonium (upper curves) for two values of the
  plasma anisotropy parameter $\xi$. The straight line corresponds to
  $T$.  Right: 1P state of bottomonium.}
\label{fig:Eb_bott}
\end{figure}
In ref.~\cite{Dumitru:2009ni} a temperature and viscosity dependent
interpolation between the short-distance potential
(\ref{eq:anisoPotlin_xi}) and $V_\infty$ has been put forward. The
bound states were obtained from the corresponding Schr\"odinger
equation~\cite{Strickland:2009ft}. The binding energies of the
charmonium and bottomonium ground states are shown in
Fig.~\ref{fig:Eb_bott} and, as expected, they decrease towards higher
$T$. As already mentioned above, this turns out to be largely due to
the decreasing continuum threshold $V_\infty(T)$. The wave function of
the $b\bar{b}$ ground state, for example, is affected little by the
medium (for $T\simle 2.5T_c$ and $\xi\simle1$) even as
$|E_{\rm{bind}}|$ drops well below $T$. The figure also shows that
$|E_{\rm{bind}}|$ increases with the anisotropy $\xi$. This can be
understood as reduced screening of both the Coulomb and the string
potentials at higher viscosity (and fixed $T$). In the future,
improved methods should be employed to determine the properties of
bound states; in particular, many-body interactions should be taken
into account by solving the Schr\"odinger equation for the
non-relativistic Green's function~\cite{Mocsy:2007yj} (incl.\
threshold effects).

The potential in an anisotropic plasma carries angular
dependence. States with non-zero angular momentum then split according
to the projection ${\bf L\cdot n}$. For the 1P state of bottomonium,
for example, the splitting is estimated to be on the order of 50~MeV;
at $T=200$~MeV, the occupation number of states with ${\bf
  L\cdot n}=0$ should be about $\exp(50/200)=1.3$ times higher than
that of states with ${\bf L\cdot n}=\pm1$.

In the static limit, the retarded and advanced HTL propagators are
real. The symmetric propagator, however, is purely imaginary. The
static potential therefore also develops an imaginary part. At short
distances, $\hat{r}\ll 1$, and including the leading viscosity
correction, it is given by~\cite{Dumitru:2009fy}
\begin{eqnarray}
i\, {\rm{Im}} ~V({\bf{r}}) =-i \frac{ g^2 C_F T}{4\pi}\,\hat{r}^2\,
\ln\frac{1}{ \hat{r}}
\left(\frac{1}{3}-\xi\frac{3-\cos(2\theta)}{20}\right)~. \label{Im_V}
\end{eqnarray}
This leads to a non-vanishing width of the bound states; for a Coulomb
wave function,
\begin{eqnarray}
\Gamma(T,\xi) &=& - \int d^3{\bf{r}}\, \left|\Psi({{r}})\right|^2 \, 
{\rm{Im}}~V({\bf{r}}) = \frac{16\pi T}{g^2
C_F}\frac{m_D^2}{M_Q^2}\left(1-\frac{\xi}{2}\right)\ln\frac{g^2C_FM_Q}{8\pi
m_D}~. \label{Gamma}
\end{eqnarray}
$\Gamma_\Upsilon$ is on the order of tens of MeV, to be compared to
the electromagnetic decay width $\Gamma_{\Upsilon\to e^+e^-} \approx
1$~keV. Hence, $\Upsilon$ states which may form in the plasma at
$T/T_C\sim 1-2$ are very hard to observe in the $e^+e^-$ channel as
the branching ratio $\Gamma_{\Upsilon\to e^+e^-}/\Gamma_\Upsilon$ is
much smaller than in vacuum. This effect would contribute to a
possible $\Upsilon\to e^+e^-$ suppression in central Au+Au collisions
at RHIC~\cite{Atomssa}.

It is interesting to compare to a strongly coupled theory. Using the
gauge-gravity duality, the static potential (or Wilson
loop)~\cite{Maldacena:1998im} and thermal effects at short
distances~\cite{Rey:1998bq} have been computed in ${\cal N}=4$
supersymmetric Yang-Mills at large (but finite~\cite{Gubser:2006qh})
t' Hooft coupling $\lambda=g^2 N$ and $N\to\infty$. At $T=0$,
\begin{equation}
V_{Q\bar{Q}}(r)= -\frac{4\pi^2}{\Gamma(1/4)^4} \frac{\sqrt{\lambda}}{r}~.
\label{maldacenaprop}
\end{equation}
The $\sim 1/r$ behavior follows from conformal invariance of the
theory. Also, the potential is non-analytic in $\lambda$. Clearly, the
coupling should not be very large or else the properties of the
resulting bound states are qualitatively different from the $\Upsilon$
etc.\ states of QCD (numerically, $4\pi^2/\Gamma(1/4)^4 \approx
0.23$).

Effects due to a hot, viscous medium may be investigated in
a theory dual to five-dimensional Gauss-Bonnet gravity which leads
to~\cite{Noronha:2009ia}
\begin{eqnarray}
V_{Q\bar{Q}}(r) &=& - \frac{2\sqrt{\lambda}}{r}
        \left( \frac{\Gamma(3/4)}{\Gamma(1/4)}\right)^2
        \left[ 1 - \frac{576\pi^2}{5}
          \frac{(rT)^4}{\eta'}
          \frac{1}{\left( 1+\eta' \right)^3}
\left( \frac{\Gamma(5/4)}{\Gamma(3/4)}\right)^4 \right]~,
\label{VQQ_lowT_etas}
\end{eqnarray}
where $\eta'\equiv\sqrt{4\pi\,\eta/s}\simge1$. The second term in the
square bracket is the leading ``thermal screening'' correction at
small $rT$.  In qualitative agreement with
eq.~(\ref{eq:anisoPotlin_xi}), the potential decreases (in magnitude)
as $T$ increases but thermal effects diminish as $\eta/s$
increases. However, note that the strong coupling
result~(\ref{VQQ_lowT_etas}) predicts a more rapid disappearance of
temperature effects as $m_Q\to \infty$; for a parametric estimate of
the thermal shift of the vacuum binding energy replace $r$ by the Bohr
radius $1/(\sqrt{\lambda} m_Q)$. The quartic dependence on $rT$
originates from the behavior of the AdS metric near the horizon.

The free energy of a single static quark (=$F_\infty/2$) in
the conformal theory dual to GB gravity is equal to~\cite{Noronha:2009ia}
\begin{equation}  \label{FQ_CFT}
F_Q = - \frac{\sqrt\lambda}{1+\eta'}\, T~.
\end{equation}
Hence, $F_Q$ decreases in magnitude with increasing viscosity. This is
qualitatively similar to the behavior in resummed perturbation
theory~\cite{Dumitru:2009ni}. Both produce pure entropy contributions
($0>F\sim T$) only and so the potential energy of the quark in the
plasma vanishes once that is removed. It will be interesting to
analyze strongly-coupled theories with broken conformal invariance
which reproduce the trace anomaly of QCD and $F_\infty$ from
eq.~(\ref{FQ_Latt}) above at $T/T_C=1-3$.


\section*{Acknowledgments}
I thank Y.~Guo, \'A.~M\'ocsy, J.~Noronha and M.~Strickland for
collaboration and RIKEN/BNL and the Dean's office, Weissman School of
Arts and Sciences, for travel support.

\end{document}